# Family of Two-Dimensional Transition Metal Dichloride's: Fundamental Properties, Structural Defects, and Environmental Stability

Andrey A. Kistanov[1,*], Stepan A. Shcherbinin[2], Romain Botella[1], Artur Davletshin[3], Wei Cao[1]

[1]Nano and Molecular Systems Research Unit, University of Oulu, Oulu 90014, Finland

[2]Peter the Great Saint Petersburg Polytechnical University, Saint Petersburg 195251, Russia

[3]Center for Subsurface Energy and the Environment, The University of Texas at Austin, Austin, Texas 78712, United States

***Corresponding Author:** andrey.kistanov@oulu.fi (AAK)

**Abstract**

A large number of novel two-dimensional (2D) materials are constantly discovered and deposed into the databases. Consolidate implementation of machine learning algorithms and density-functional theory (DFT)-based predictions have allowed creating several databases containing an unimaginable amount of 2D samples. The next step in this chain, the investigation leads to a comprehensive study of the functionality of the invented materials. In this work, a family of transition metal dichlorides has been screened out for systematical investigation of their structural stability, fundamental properties, structural defects, and environmental stability via DFT-based calculations. The work highlights the importance of using the potential of the invented materials and proposes a comprehensive characterization of a new family of 2D materials.

**KEYWORDS**: DFT, 2D material, point defect, environmental stability, metal chlorine

    More than a decade has passed since the era of 2D materials' has begun and the research effort around their discovery and applications continues unabated.[1-4]. This large family of materials presents unique properties, ranging from electronic to mechanical,[5-7] which largely account for the high research activity of the field. Fueling the rise of 2D materials, their prediction and discovery using computational methods are revealing their wide diversity, both structural and compositional.[8-10] Prediction strategies are of two types: combinatorial and top-down. Combinatorial approaches are based on combining an atomic composition and a crystal structure to obtain previously unexplored 2D atomic structures,[11] while top-down approaches focus on slicing bulk materials into mono to few-layers assemblies.[12] These methodologies are often upscaled to high-throughput systems predicting many 2D materials, which constitutes a great achievement towards the full exploration of this part of the material space. Several imposing databases exist, filled with the results of such endeavors.[13-15] Despite these databases being valuable warehouses of 2D materials, they can still be further completed with newly discovered ones.[16] Moreover, the enormous amount of already existing 2D candidates lacks specificity towards prospective applications.

    A promising path to identify an application protential of the 2D species from the said databases is to study an individual family of 2D compounds with similar chemical forms.[17] For

instance, investigation of specifics of structure and properties of a theoretically designed family of transition-metal diboride has helped to find application for them in the conversion of $CO_2$.[18] The development of transition metal carbides and nitrides has allowed selecting $Ti_3C_2T_x$ monolayers possessing the highest effective Young's modulus of ~0.33 TPa among other solution-processed 2D materials, including graphene oxide.[19] The criteria to pick 2D materials for most of the known applications are already well understood,[20] with one of the most important being the environmental stability, tunability of electronic structure, and mechanical strength. An even better criterion would be the commercial availability or/and well-developed synthesis process of 2D materials, lifting technical locks impairing the investigations towards their application.

Van der Waals layered transition metal dichlorides ($MCl_2$) are starting to be available,[21] while being found in databases.[13] Therefore, they constitute a very good candidates for more in-depth studies. Metal halides are commonly investigated in perovskite structures for several applications going from light-emitting devices[22,23] to nanospintronics,[24] and show tunable properties when shrinking from bulk layered material to monolayer.[23] While individual transition metal halides have been studied for their unique magnetic properties,[25,26] while their environmental stability and electronic and mechanical properties have been seldom studied so far.

The present work is dedicated to a DFT simulation-based systematic search of all possible existing materials in a family of 2D $MCl_2$. Their structural and thermodynamical stabilities are determined by means of phonon dispersion analysis and *ab initio* molecular dynamic (AIMD) simulations. The characteristic features of screened-out 2D $MCl_2$ are further analyzed to obtain comprehensive knowledge on their electronic and mechanical properties. Point defects formation and surface activity of the 2D $MCl_2$ towards environmental molecules are considered to facilitate their experimental observation and enlarge the area of their possible applications.

The unit cell structure of monolayer of $MCl_2$ (Figure 1) was designed based on the geometry of primitive unit cell of a monolayer of trigonal $FeCl_2$ available in the 2DMatPedia database (ID dm-3574)[13] and all transition metals (according to the Periodic Table) were considered as M element. For the unit cell of each designed structure a geometry optimization was performed. The structural stability of those optimized structures was verified by calculating phonon dispersion spectra while their thermodynamic stability was controlled by *ab initio* molecular dynamics (AIMD) calculations.[27] Based on those simulations a stable modification of 2D $MCl_2$ was proposed.

The top and side views of the unitcell of 2D $MCl_2$ are shown in Figure 1a. The unitcell of 2D $MCl_2$ consists of one atom of transition metal and two atoms of chlorine. 2D $MCl_2$ possess a trigonal lattice with the space group 164 *P-3m1*. The kinetic stability of all possible 2D $MCl_2$ is considered by calculating the phonon dispersion spectra along the high symmetry directions ($\Gamma \rightarrow M \rightarrow K \rightarrow \Gamma$) of the Brillouin zone. Among all 2D $MCl_2$ only 2D $FeCl_2$, 2D $CdCl_2$, 2D $MnCl_2$, 2D $NiCl_2$, 2D $VCl_2$, and 2D $ZnCl_2$ are found to be stable, as their phonon dispersion curves are positive in whole Brillouin zone and the transverse acoustic (TA), longitudinal acoustic (LA), and out-of-plane acoustic (ZA) modes of these materials display the normal linear dispersion around the $\Gamma$ point (Figure 2). Therefore, only these 2D $MCl_2$ are further considered in this study. According to AIMD simulations conducted, the listed materials are also show thermal stability at 300 K (Figure S1). The structural parameters of all stable 2D $MCl_2$ are collected in Table S1.

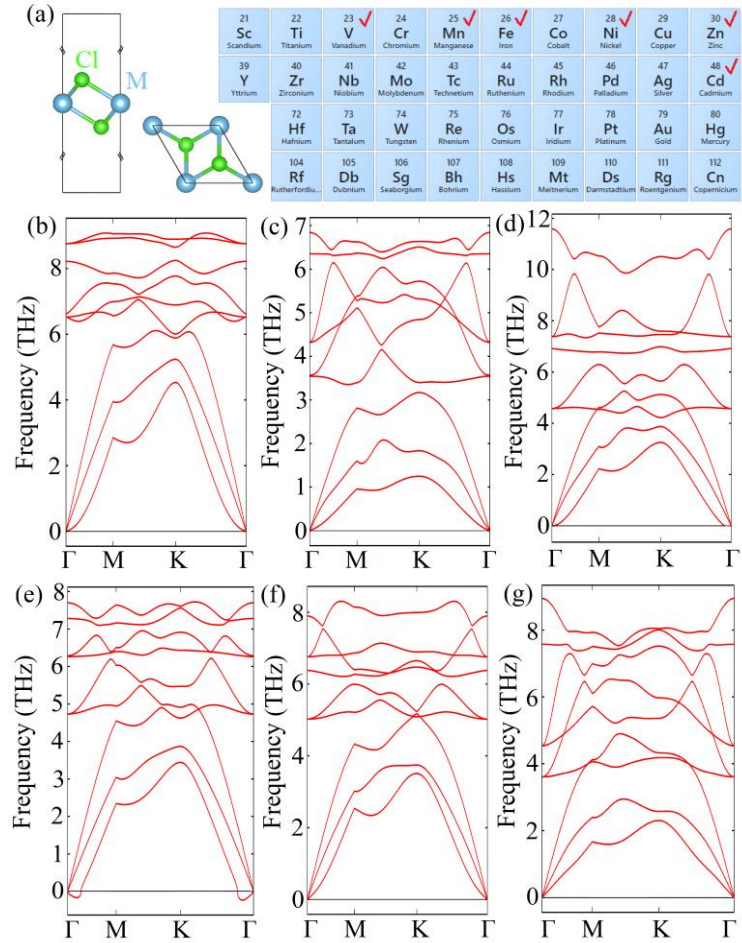

**Figure 1.** (a) The unit cell structure of 2D MCl$_2$ and a part of the Periodic Table with the marked range of the screened M elements. Phonon dispersion curves for (b) 2D FeCl$_2$, (c) 2D CdCl$_2$, (d) 2D MnCl$_2$, (e) 2D NiCl$_2$, (f) 2D VCl$_2$, (g) 2D ZnCl$_2$.

To evaluate possible applications of the above 2D MCl$_2$ in electronic and straintronic devices, their electronic and mechanical properties are further considered. The band structure of 2D MCl$_2$ obtained using both the Perdew–Burke–Ernzerhof (PBE) exchange-correlation functional under the generalized gradient approximation (GGA) and the Heyd–Scuseria–Ernzerhof (HSE06) functional is plotted in Figure 2, while the partial density of states (PDOS) calculated using GGA PBE approach are plotted in Figure S2. It should be noted that the discrepancy of the band gap sizes calculated via the PBE GGA and HSE06 methods is due to lower accuracy of the PBE GGA approach which often underestimates the width of band gap.[28] Therefore, the HSE06 method is expected to provide a better match with experimental results. The band gap values $E_g$ calculated for 2D MCl$_2$ are collected in Table 1. The HSE06 approach predicts 2D FeCl$_2$ is a direct band gap semiconductor with $E_g$ of 4.10 eV (0.85 eV according to PBE GGA). The conduction band minimum (CBM) and valence band maximum (VBM) are located between the Γ and K points. According to the PDOS plot in Figure S2a, the CBM forms because of a strong mixing of Cl $p$-states and Fe $d$-states and the VBM mainly consists of Fe $d$-states. 2D CdCl$_2$ is found to be a direct band gap semiconductor with $E_g$ of 4.88 eV (3.40 eV according to PBE GGA) and CBM and VBM located at the Γ point (Figure 2b). PDOS plot in Figure S2b shows the CB and VB of 2D CdCl$_2$ are mainly formed by Cl $p$-states. An indirect $E_g$ of 4.76 eV (1.60 eV according to PBE GGA) is found for the 2D MnCl$_2$ (Figure 2c). The CBM is located between the Γ and K points and consists of Mn $d$-states, while the VBM is located at the Γ point and forms because of a strong mixing of Cl $p$-states and Mn $d$-states (Figure S2c). For 2D NiCl$_2$ (Figure 2d) an indirect $E_g$ of 4.10 eV (1.02 eV according to PBE GGA) is predicted. The CBM is located between the Γ and K points, while the VBM is located at the Γ point and both CB and VB are formed because of a strong mixing of Cl $p$-states and Ni $d$-states (Figure S2d). Figure 2e shows 2D VCl$_2$ is a direct band gap semiconductor with $E_g$ of 3.21 eV (0.45 eV according to PBE GGA). Both CBM and VBM are located in the vicinity of the K points. The CB forms because of a strong mixing of Cl $p$-states and V $d$-states, while the VB only consists of V $d$-states (Figure S2e). An indirect $E_g$ of 6.14 eV (4.52 eV according to PBE GGA) is found for 2D ZnCl$_2$ Figure 2f. The CBM is located in the vicinity of the K point and CB consists of only Cl $p$-states, while the VBM is located in the vicinity of the Γ point and VB is formed by Cl $p$-states and Zn $d$-states (Figure S2f).

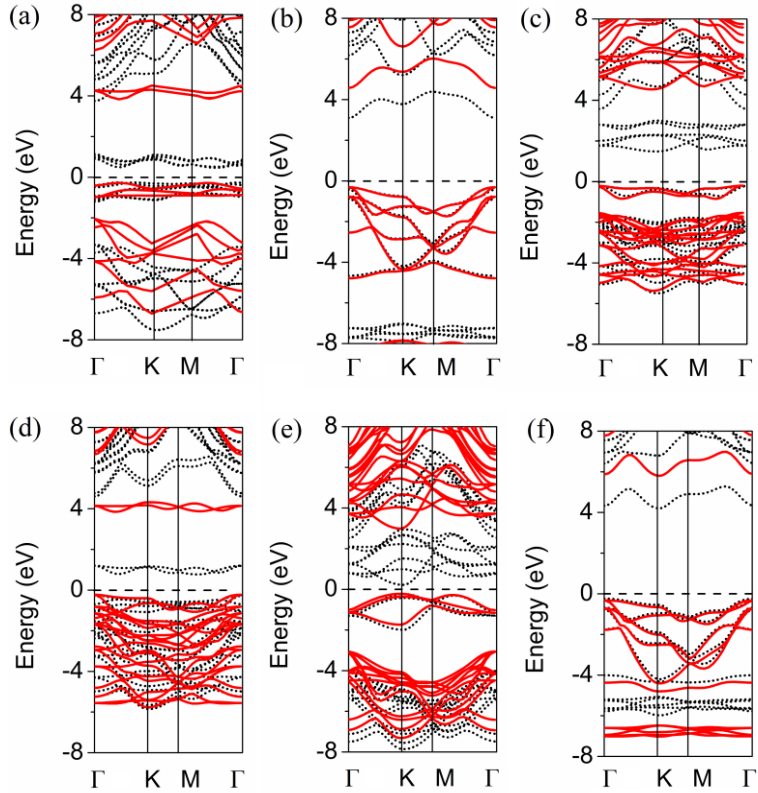

**Figure 2.** The band structure of (a) 2D FeCl$_2$, (b) 2D CdCl$_2$, (c) 2D MnCl$_2$, (d) 2D NiCl$_2$, (e) 2D VCl$_2$, and (f) 2D ZnCl$_2$. The black and red lines show the band structure calculated by the PBE and HSE approaches.

Table 1 also contains work function (WF) values for studied 2D MCl$_2$. 2D FeCl$_2$ and 2D VCl$_2$ possesses relatively low WF of 4.66 eV and 3.90 eV, respectively. In turns, 2D CdCl$_2$, 2D MnCl$_2$, 2D NiCl$_2$, and 2D ZnCl$_2$ have high WF values of 7.09 eV, 6.15 eV, 6.32 eV, and 7.26 eV, respectively, that are higher than these of most 2D materials[29] such as, graphene (4.60 eV) and borophene (5.31 eV), and bulk metals,[30] such as Ni (5.23 eV) and Pt (5.65 eV). The relatively low work function of 2D FeCl$_2$ and 2D VCl$_2$ can be attributed to the nature of Cl atomic states around the Fermi level consisting of the out-of-plane $p_z$ states (Figure S3a), which lie above the in-plane s–p hybridized states. As a result, the ionization of 2D FeCl$_2$ and 2D VCl$_2$ is comparable to that of graphene, while in 2D CdCl$_2$, 2D MnCl$_2$, 2D NiCl$_2$, and 2D ZnCl$_2$, in-plane $p_x$ and $p_y$ states of Cl are predominant in the vicinity of the Fermi level (Figure S3b) which explains their high WF values.

The calculated spatial dependencies of Young's modulus, shear modulus, and Poisson's ratio of 2D FeCl$_2$ are presented in Figure 3. It is seen that these quantities are direction-independent. A similar isotropic distribution of Young's modulus, shear modulus, and Poisson's ratio is found for all considered 2D MCl$_2$ (Figure S4). Therefore, each considered 2D MCl$_2$ can be characterized by in plane Young's modulus, shear modulus, and Poisson's ratio. Among all considered 2D MCl$_2$, 2D FeCl$_2$ and 2D NiCl$_2$ possess the highest Young's modulus of 110 GPa and of 107 GPa and shear modulus of 45 GPa and of 43 GPa, respectively (Table 1), which are lower than those of graphene[31] and MoS$_2$.[32] Importantly, that Poisson's ratio of the considered materials fall in the range from 0 to 0.5 (Table 1), showing their high elasticity in a line with other 2D materials.[33]

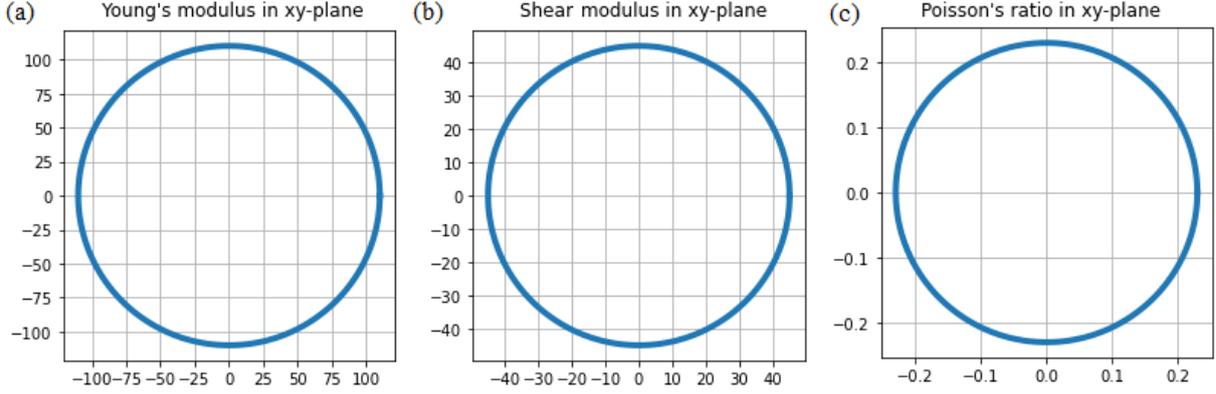

**Figure 3.** Spatial dependencies of (a) Young's modulus (in GPa), (b) shear modulus (in GPa), and (c) Poisson's ratio for 2D FeCl$_2$.

**Table 1.** Band gap size $E_g$ (HSE method), workfunction WF, Young's modulus $E$, Shear modulus $G$, and Poisson's ratio $v$ of considered 2D MCl$_2$.

|  | $E_g$, eV | WF, eV | $E$, GPa | $G$, GPa | $v$ |
|---|---|---|---|---|---|
| 2D FeCl$_2$ | 4.10 (direct) | 4.66 | 110 | 45 | 0.23 |
| 2D CdCl$_2$ | 4.88 (direct) | 7.09 | 36 | 13 | 0.38 |
| 2D MnCl$_2$ | 4.76 (indirect) | 6.15 | 47 | 18 | 0.32 |
| 2D NiCl$_2$ | 4.10 (indirect) | 6.32 | 107 | 43 | 0.24 |
| 2D VCl$_2$ | 3.21 (direct) | 3.90 | 83 | 33 | 0.29 |
| 2D ZnCl$_2$ | 6.14 (indirect) | 7.26 | 46 | 16 | 0.46 |

2D materials commonly host structural defects such as point defects,[34,35] which are formed spontaneously in real systems, while their type and concentration can certainly be controlled by ion/electron irradiation or by mechanical damage of the material's surface.[36] Such defects may change the local structure of 2D materials and influence their properties.[37] Therefore, a comprehensive study on the formation typical point defects in MCl$_2$ is further conducted. Figure 4 (left panels) shows the atomic structure of 2D FeCl$_2$ and a geometry of the most common point defects found to be stable for this structure. The stability of point defects in 2D $MCl_2$ is considered in terms of their formation energy $E_{form}$. Similarly, an atomic structure of other studied 2D $MCl_2$ and a geometry of the most common point defects stable in these structures are shown in Figures S5-S9 (left panels). These are a single Cl vacancy ($SV_{Cl}$), single M vacancy ($SV_M$), double Cl vacancy with one Cl atom on each side of the layer ($DV^I_{Cl}$), double Cl vacancy on the same side of the layer ($DV^{II}_{Cl}$), and double vacancy of one Cl atom and one M atom ($DV_{MCl}$). It should be noted that a TMD-like structures, as it is in case of 2D $MCl_2$, does not contain SW defects.[38] Two SV defects can be introduced into 2D $MCl_2$ by removing the M or Cl atom from its surface, as shown in Figures 4b and c (left panels), respectively. 2D $MCl_2$ can also host three various DV defects. The first is the $DV^I_{Cl}$ defect which is created by removing one Cl atom from one side of the 2D $MCl_2$ layer and one Cl atom from another side of the 2D $MCl_2$ layer (Figure 4d, left panels). The $DV^{II}_{Cl}$ defect is created by removing two neighbouring Cl atoms from one side of the 2D $MCl_2$ layer (Figure 4e, left panels). The remaining $DV_{MCl}$ defect is formed when the neighbouring M atom and Cl atom are removed from the 2D $MCl_2$ layer (Figure 4f, left panels).

The calculated $E_{form}$ of the considered defects in 2D $MCl_2$ is presented in Table S2. According to Table S2, the $SV_{Cl}$ defect has the lowest $E_{form}$ in all the considered 2D $MCl_2$. In 2D $FeCl_2$ $E_{form}$ of the $SV_{Cl}$ defect is as low as 1.04 eV, which is comparable to $E_{form}$ of SV in phosphorene (~1–2 eV)[39] and ~2 times lower than that of SV in the most common 2D TMD material, $MoS_2$ (~2.12 eV).[40] Therefore, a low $E_{form}$ of $SV_{Cl}$ defect in 2D $FeCl_2$ may lead to its instability at room temperature, similarly to the case of phosphorene.[37] Despite a low $E_{form}$ the $SV_{Cl}$ defect in 2D $FeCl_2$ possesses high stability which is confirmed by AIMD simulations at room temperature for 3 ps (movie 1 in SI). $E_{form}$ of $SV_{Cl}$ of 3.23 eV in 2D $NiCl_2$ is higher than that of the SV defect in $MoS_2$, while still significantly lower $E_{form}$ of SV in graphene (7.5 eV).[41] For 2D $CdCl_2$, 2D $MnCl_2$, 2D $VCl_2$, and 2D $ZnCl_2$ $E_{form}$ of the $SV_{Cl}$ defect are 4.75 eV, 4.56 eV, 5.14 eV, and 5.02 eV, respectively, that are significantly higher than that of the SV defect in $MoS_2$, but still lower than that of the SV defect in graphene. It should be noted, that DV defects in 2D $MCl_2$ (except 2D $FeCl_2$) has higher $E_{form}$ (~7-10 eV) than that of DV defects in most common 2D materials including graphene (~8 eV)[41] and $MoS_2$ (~4 eV).[42]

A remarkable difference in $E_{form}$ of SV defects in 2D $MCl_2$ can be attributed to the difference in the electronegativity of M elements compared to that of Cl.[43] It is known that, if the difference in electronegativity of a bonded metal and non-metal is greater than ~1.5, a compound is expected to be ionic, while a covalent type of bonding is expected when the electronegativity of a bonded metal and non-metal is less than ~1.5. Therefore, the bonds in 2D $FeCl_2$ and 2D $NiCl_2$ are expected to have covalent nature, as the difference in the electronegativity of Cl (3.0) and both Fe (1.8) and Ni (1.9) is greater than ~1.5. On the other hand, the difference in the electronegativity of Cl (3.0) and both Cd (1.7), Mn (1.5), V (1.6), and Zn (1.6) is close to ~1.5, which suggest the existence of ionic bonds between these compounds. To support this conclusion, electron localization function for 2D $FeCl_2$ and 2D $ZnCl_2$ is analyzed.[44-47] In the case of 2D $FeCl_2$ (Figure S5a) the electron localization observed on Fe atoms and partially on the Fe-Cl bond, which confirm the existence of an ionocovalent type of bonding in 2D $FeCl_2$. In the case of 2D $ZnCl_2$ (Figure S5b), the electron localization basin is spherical and completely migrates to the Zn atom so that basins are all surrounding the respective cores, suggesting an ionic bond in 2D $ZnCl_2$. Therefore, strong ionic bonds in 2D $CdCl_2$, 2D $MnCl_2$, 2D $VCl_2$, and 2D $ZnCl_2$ can explain their high stability against the formation of most point defects compared to 2D $FeCl_2$, 2D $NiCl_2$, and common 2D materials.

To facilitate experimental identification of point defects in 2D $MCl_2$ the simulated scanning tunnelling microscopy (STM) images are obtained for perfect and defect-containing 2D $MCl_2$. A constant height mode characterization method is used in all cases. The STM images of the perfect and defect-containing 2D $FeCl_2$ are presented in Figure 4 (right panels), while the STM images of defects for other studied 2D $MCl_2$ are shown in SI (Figures S6-S10). Defects are easy to recognize at STM images and are well correlate with their atomic structures. For instance, the STM image at Figure 4b (right panel) clearly reflects the SV$_{Cl}$ defect with a triangle formed of three bright spots characterizing three Cl atoms inside of which one Cl is missing. Similarly, Figure 4c (right panel), the SV$_{Fe}$ defect presented by a triangle formed of three big bright spots characterizing three Cl atoms and a pentagon formed of five small bright spots characterizing five Fe atoms inside of which one Fe is missing. The most complicated task is to differentiate the DV$^I_{Cl}$ defect, which may be confused with the SV$_{Cl}$ defect. However, opposed to the SV$_{Cl}$ defect, in the case of DV$^I_{Cl}$, four small bright spots in a form of a parallelogram reflecting four Fe atoms with one missing Cl atom inside are clearly visible (Figure 4d, right panel). The STM image of the DV$^{II}_{Cl}$ defect is presented in Figure 4e (right panel), there the formation of the triangle of three small bright spots as three Fe atoms are shifted due to the absence of two neighboring Cl atoms (two dim spots are absent) is seen. The DV$_{FeCl}$ defect visible of the STM image (Figure 4f, right panel) as there is one small (Fe atom) and one big (Cl atom) bright spots are clearly missing.

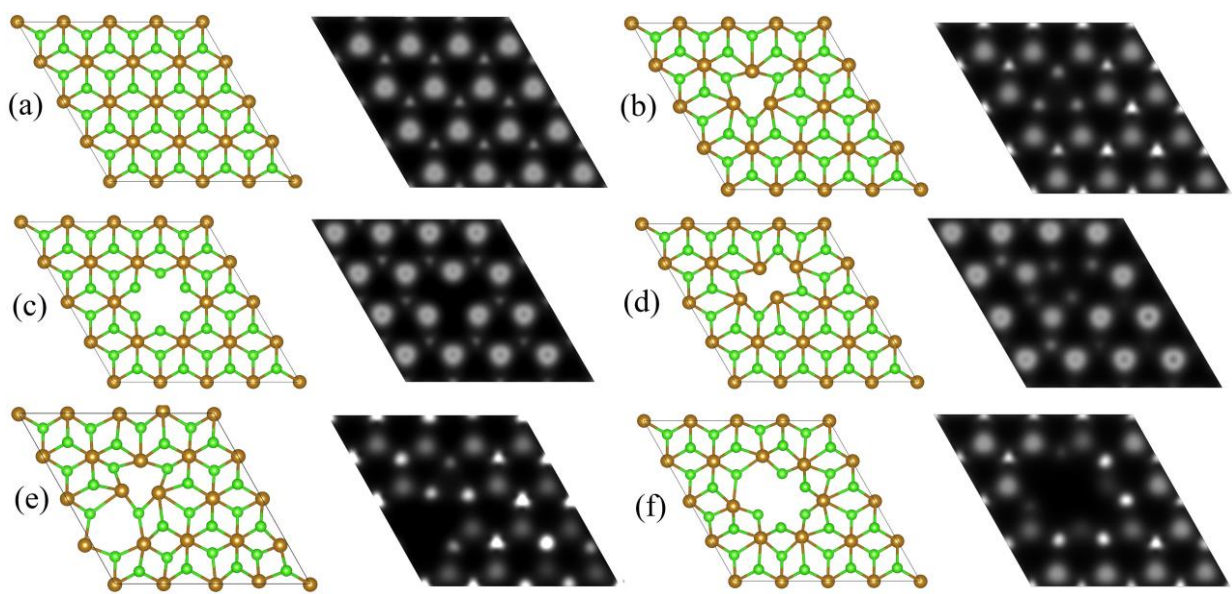

**Figure 4.** Atomic structure (the left panels) and STM images at a constant height mode (the right panels) of (a) pure, (b) SV$_{Cl}$-containing, (c) SV$_M$-containing, (d) DV$^I_{Cl}$-containing, (e) DV$^{II}_{Cl}$-containing, and (f) DV$_{MCl}$-containing 2D FeCl$_2$.

It is well known that 2D materials possess high sensitivity to the environmental conditions.[48-50] To determine the behaviour of 2D MCl$_2$ under environmental conditions, particularly in the presence of moisture, their interaction with H$_2$O and O$_2$ molecules is considered. All possible adsorbing configurations of H$_2$O and O$_2$ on studied 2D MCl$_2$ are considered. The determined lowest-energy configuration, together with adsorption energy $E_a$ for the H$_2$O and O$_2$ molecules on 2D MCl$_2$, are presented in Figures S11 and S12. In cases of 2D FeCl$_2$, 2D CdCl$_2$, 2D MnCl$_2$, 2D NiCl$_2$, and 2D ZnCl$_2$, the H$_2$O and O$_2$ molecules are located at above the metal site with the O atom directed to the surface. $E_a$ of H$_2$O and O$_2$ on these materials is comparably high (Figures S11 and S12) and is comparable to that of H$_2$O and O$_2$ on other common 2D materials (Table S3), such as graphene,[51] 2D pnictogens,[49,52] and family of 2D phosphorus carbides.[53] 2D VCl$_2$ stands out of its counterparts as the H$_2$O and O$_2$ molecules have 2 times lower $E_a$ on its surface and is located the Cl atom with both H atoms (in case of H$_2$O) and O atom (in case of O$_2$) tending to two other Cl atoms at the surface. The calculated $E_a$ of H$_2$O and O$_2$ on 2D MCl$_2$ is comparably high, therefore, these materials are supposed to be environmentally stable. This is also confirmed via AIMD simulations where a weak interaction of the 2D FeCl$_2$ surfaces with H$_2$O (movie 2 in SI) and O$_2$ (movie 3 in SI) at room temperature is shown. It should be noted that metal chlorines usually possess strong electron donating/accepting abilities, making these materials active for adsorbents.[54] One of the reasons for that can be their constituent elements with weak/strong electronegativities or high ionicity. Another reason can be a comparably low $E_{form}$ of defects in 2D MCl$_2$, which can also affect their stability. For instance, it is found that $E_a$ of H$_2$O on 2D MCl$_2$ lowers by 6 times (from -0.12 eV to -0.66 eV) in the presence of SV$_{Cl}$ defect compared to that of H$_2$O on pure 2D MCl$_2$. On the other hand, as it has been shown for metal (hydr)oxides, the adsorption of various species on metal chloride surfaces under moisture and/or water saturated conditions can be hindered.[55] Therefore, oxygen-passivated and water saturated metal-containing materials can exhibit higher stability to adsorbents. It can be concluded that despite studied 2D MCl$_2$ are found to be stable under environmental conditions, their stability may be affected by many factors, such as surface hydration and defect formation.

In summary, in this work following a sequential search over existing 2D materials databases and the subsequent systematic screening of possible atomic combinations, a new family of 2D MCl$_2$, consisting of 2D FeCl$_2$, 2D CdCl$_2$, 2D MnCl$_2$, 2D NiCl$_2$, 2D VCl$_2$, and 2D ZnCl$_2$, has been identified. DFT-based simulation has been implemented to prove the structural stability of the screened-out materials and systematically study their fundamental properties and structural changes under certain conditions, such as presence of point defects and moisture environment. It is shown that 2D MCl$_2$'s, due to their electronic and mechanical properties versatile candidacies in the in the semiconductor industry, while the defect-related and ambient stabilities demonstrate their durability and manipulation feasibilities. Particularly, 2D MnCl$_2$, 2D NiCl$_2$, and 2D ZnCl$_2$ due to their high WF values can be used in carrier transport nanoelectronic devices, while high Young's modulus and shear modulus of 2D FeCl$_2$ and 2D NiCl$_2$ make them good candidates for straintronic devices.[56] The work highlighted the importance of the developing 2D material's databases and the need for a deep investigation and characterization of materials available in the existing databases.

**Computational Methods**

All calculations were performed using the plane-wave method as implemented in the Vienna Ab initio Simulation Package (VASP).[57] The PBE exchange−correlation functional under the GGA[58] was used for the geometry optimization calculations while the electronic structure calculations were supplemented with the HSE functional.[59] The considered supercells of 2D MCl$_2$ were composed of 4×4×1 unit cells (16 M and 32 Cl atoms) to avoid non-physical interactions between periodic images while keeping affordable computational demand. The optimization was stopped once the atomic forces and total energy values were smaller than $10^{-4}$ eV/Å and $10^{-8}$ eV. The first Brillouin zone was sampled with a 15×15×1 k-mesh grid for the unit cell and 3×3×1 k-mesh grid for the 4×4×1 supercell. The kinetic energy cut-off of 520 eV was set. The periodic boundary conditions were applied for the two in-plane transverse directions while a vacuum space of 20 Å was introduced to the direction perpendicular to the surface plane. Under such conditions, concentration of SVs and DVs defects was 2.08% (one M/Cl atom per 48 atoms) and 4.17% (two M/Cl atom per 48 atoms), respectively.

The finite displacement approach as implemented in the Phonopy code[60] was used to simulate phonon dispersion spectra. AIMD simulation lasts for 5 ps with a time step of 1.0 fs and the temperature of 300 K was controlled by a Nose–Hoover thermostat. STM images were simulated via the Tersoff-Hamann approach.[61]

The stress-strain relation was used to calculate components of the stiffness matrix for the considered structures.[62] For these calculations approximate values of interlayer distances was used. The interlayer distance was considered to be such a distance at which the force of action between the layers becomes less than 0.01 eV/A. Based on obtained stiffness matrix the Young's modulus, shear modulus, and Poisson's ratio were calculated and directional dependencies of these quantities were defined using ELATE software for analysis of elastic tensors.[63]

The stability of the considered point defects in 2D $MCl_2$ was considered based on their formation energy $E_{form}$, which was calculated as

$$E_{form} = E_{defect} - E_{perfect} + N_M E_M + N_{Cl} E_{Cl} \qquad (1)$$

where $E_{defect}$ and $E_{perfect}$ are the total energies of perfect and defect-containing 2D $MCl_2$, $E_M$ and $E_{Cl}$ are the energies of a single transition metal and chlorine atom, and $N_M$ and $N_{Cl}$ correspond to the number of the removed transition metal and chlorine atoms.

## ACKNOWLEDGEMENT

A.A.K., R.B. and W.C. acknowledge funding provided by the European Research Council (ERC) under the European Union's Horizon 2020 research and innovation programme (grant agreement No. 101002219). S.A.Sh thanks funding provided by the Ministry of Science and Higher Education of the Russian Federation as part of World-class Research Center program: Advanced Digital Technologies (contract No. 075-15-2020-934 dated 17.11.2020). A.D. thanks funding provided by the American Chemical Society. The authors also acknowledge CSC–IT Center for Science, Finland and Peter the Great Saint-Petersburg Polytechnic University Supercomputer Center "Polytechnic" for computational resources.

## REFERENCES

(1) Zhou, S.; Zhao, J. Photoinduced spin injection and ferromagnetism in 2D group III monochalcogenides. *J. Phys. Chem. Lett.* **2022**, *13*(2), 590–597.

(2) Kistanov, A. A.; Shcherbinin, S. A.; Ustiuzhanina, S. V.; Huttula, M.; Cao, W.; Nikitenko, V. R.; Prezhdo, O. V. First-principles Prediction of two-dimensional $B_3C_2P_3$ and $B_2C_4P_2$: Structural stability, fundamental properties, and renewable energy applications. *J. Phys. Chem. Lett.* **2021**, *12*(13), 3436–3442.

(3) Miro, P.; Audiffred, M.; Heine, T. An atlas of two-dimensional materials. *Chem. Soc. Rev.* **2014**, *43*, 6537–6554.

(4) Bergeron, H.; Lebedev, D.; Hersam, M. C. Polymorphism in post-dichalcogenide two-dimensional materials. *Chem. Rev.* **2021**, *121*, 2713–2775.

(5) Li, F.; Zhang, X.; Fu, Y.; Wang, Y.; Bergara, A.; Yang, G. Ba with unusual oxidation states in Ba chalcogenides under pressure. *J. Phys. Chem. Lett.* **2021**, *12*(35), 8481–8488.

(6) Kochaev, A.; Katin, K.; Maslov, M.; Meftakhutdinov, R. AA-stacked borophene-graphene bilayer with covalent bonding: Ab Initio investigation of structural, electronic and elastic properties. *J. Phys. Chem. Lett.* **2020**, *11*(14), 5668–5673.

(7) Roldan, R.; Chirolli, L.; Prada, E.; Silva-Guillen, J., A.; San-Jose, Guinea, F. Theory of 2D crystals: Graphene and beyond. *Chem. Soc. Rev.* **2017**, *46*, 4387–4399.

(8) Prezhdo, O. V. Advancing physical chemistry with machine learning. *J. Phys. Chem. Lett.* **2020**, *11*, 9656–9658.

(9) Springer, M., A.; Liu, T.-J.; Kuc, A.; Heine, T. Topological two-dimensional polymers. *Chem. Soc. Rev.* **2020**, *49*, 2007–2019.

(10) Yan, L.; Silveira, O. J.; Alldritt, B.; Krejci, O.; Foster, A. S.; Liljeroth, P. Synthesis and local probe gating of a monolayer metal-organic framework. *Adv. Funct. Mater.* **2021**, *31*(22), 2100519.

(11) Gu, T.; Luo, W.; Xiang, H. Prediction of two-dimensional materials by the global optimization approach. *WIREs Comput. Mol. Sci.* **2017**, *7*(2), e1295.

(12) Mounet, N.; Gibertini, M.; Schwaller, P.; Campi, D.; Merkys, A.; Marrazzo, A.; Sohier, T.; Castelli, I. E.; Cepellotti, A.; Pizzi, G.; Marzari, N. Two-dimensional materials from high-throughput computational exfoliation of experimentally known compounds. *Nat. Nanotech.* **2018**, *13*, 246–252.

(13) Zhou, J.; Shen, L.; Costa, M.D.; Ong, S. P.; Huck, P.; Lu, Y.; Ma, X.; Chen, Y.; Tang, H.; Feng, Y. P. 2DMatPedia, an open computational database of two-dimensional materials from top-down and bottom-up approaches. *Sci. Data* **2019**, *6*, 86.

(14) Zhang, T.; Jiang, Y.; Song, Z.; Huang, H.; He, Y.; Fang, Z.; Weng, H.; Fang, C. Catalogue of topological electronic materials. *Nature* **2019**, *566*, 475–479.

(15) Rasmussen, F. A.; Thygesen, K. S. Computational 2D materials database: Electronic structure of transition-metal dichalcogenides and oxides. *J. Phys. Chem. C* **2015**, *119*, 13169–13183.

(16) Momeni, K.; Ji, Y.; Wang, Y.; Paul, S.; Neshani, S.; Yilmaz, D. E.; Shin, Y. K.; Zhang, D.; Jiang, J. W.; Park, H. S.; Sinnott, S.; van Duin, A.; Crespi, V.; Chen, L. Q. Multiscale computational understanding and growth of 2D materials: A review. *npj Comput. Mater.* **2020**, *6*, 22.

(17) Zhao, J.; Zhao, Y.; He, H.; Zhou, P.; Liang, Y.; Frauenheim, T. Stacking engineering: A boosting strategy for 2D photocatalysts. *J. Phys. Chem. Lett.* **2021**, *12*(41), 10190–10196.

(18) Yuan, H.; Li, Z.; Yang, J. Transition-metal diboride: A new family of two-dimensional materials designed for selective $CO_2$ electroreduction. *J. Phys. Chem. C* **2019**, *123*(26), 16294–16299.

(19) Lipatov, A.; Lu, H.; Alhabeb, M.; Anasori, B.; Gruverman, A.; Gogotsi, Y.; Sinitskii, A. Elastic properties of 2D $Ti_3C_2T_x$ Mxene monolayers and bilayers. *Sci. Adv.* **2018**, *4*(6), eaat0491.

(20) Singh, A. K.; Mathew, K.; Zhuang, H. L.; Hennig, R. G. Computational screening of 2D materials for photocatalysis. *J. Phys. Chem. Lett.* **2015**, *6*, 1087–1098.

(21) Cai, S.; Yang, F.; Gao, C. FeCl$_2$ monolayer on HOPG: Art of growth and momentum filtering effect. *Nanoscale* **2020**, *12*, 16041–16045.

(22) Huang, H.; Weng, B.; Zhang, H.; Lai, F.; Long, J.; Hofkens, J.; Douthwaite, R. E.; Steele, J. A.; Roeffaers, M. B. J. Solar-to-chemical fuel conversion via metal halide perovskite solar-driven electrocatalysis. *J. Phys. Chem. Lett.* **2022**, *13*(1), 25–41.

(23) Weidman, M. C.; Goodman, A. J.; Tisdale, W. A. Colloidal halide perovskite nanoplatelets: An exciting new class of semiconductor nanomaterials. *Chem. Mat.* **2017**, *29*, 5019–5030.

(24) Kulish, V. V.; Huang, W. Single-layer metal halides MX$_2$ (X = Cl, Br, I): Stability and tunable magnetism from first principles and Monte Carlo simulations. *J. Mater. Chem. C* **2017**, *5*, 8734–8741.

(25) McGuire, M. A. Crystal and magnetic structures in layered, transition metal dihalides and trihalides. *Crystals* **2017**, *7*, 121.

(26) Kezilebieke, S.; Silveira, O. J.; Huda, M. N.; Vano, V.; Aapro, M.; Ganguli, S. C.; Lahtinen, J.; Mansell, R.; van Dijken, S.; Foster, A. S.; Liljeroth, P. Electronic and magnetic characterization of epitaxial CrBr$_3$ monolayers on a superconducting substrate. *Adv. Mater.* **2021**, *33*(23), 2006850.

(27) Henkelman, G.; Uberuaga, B. P.; Jonsson, H. A Climbing image nudged elastic band method for finding saddle points and minimum energy paths. *J. Chem. Phys.* **2000**, *113*, 9901.

(28) Perdew, J. P.; Kurth, S.; Zupan, A.; Blaha, P. Accurate density functional with correct formal properties: A step beyond the generalized gradient approximation. *Phys. Rev. Lett.* **1999**, *82*, 2544.

(29) Kistanov, A. A.; Cai, Y.; Zhou, K.; Srikanth, N.; Dmitriev, S. V.; Zhang, Y.-W. Exploring the charge localization and band gap opening of borophene: A first-principles study. *Nanoscale*, **2018**, *10*(3), 1403–1410.

(30) Lee, E. J. H.; Balasubramanian, K.; Weitz, R. T.; Burghard, M.; Kern, K. Contact and edge effects in graphene devices. *Nat. Nanotechnol.* **2008**, *3*, 486–490.

(31) Cao, K.; Feng, S.; Han, Y.; Gao, L.; Ly, T. H.; Xu, Z.; Lu Y. Elastic straining of free-standing monolayer graphene. *Nat. Commun.* **2020**, *11*, 284.

(32) Li, Y.; Yu, C.; Gan, Y.; Jiang, P.; Yu, J.; Ou, Y.; Zou, D. F.; Huang, C.; Wang, J.; Jia, T.; Luo, Q.; Yu, X. F.; Zhao, H.; Gao, C. F., Li J. Mapping the elastic properties of two-dimensional mos$_2$ via bimodal atomic force microscopy and finite element simulation. *npj Comput. Mater.* **2018**, *4*, 49.

(33) Zhang, C.; He, T.; Matta, S. K.; Liao, T.; Kou, L.; Chen, Z.; Du, A. Predicting novel 2D MB$_2$ (M = Ti, Hf, V, Nb, Ta) monolayers with ultrafast dirac transport channel and electron-orbital controlled negative poisson's ratio. *J. Phys. Chem. Lett.* **2019**, *10*(10), 2567–2573.

(34) Zhang, L.; Chu, W.; Zheng, Q.; Benderskii, A. V.; Prezhdo, O. V.; Zhao, J. Suppression of electron-hole recombination by intrinsic defects in 2D monoelemental material. *J. Phys. Chem. Lett.* **2019**, *10*(20), 6151–6158.

(35) Zhang, L.; Vasenko, A. S.; Zhao, J.; Prezhdo, O. V. Mono-elemental properties of 2D black phosphorus ensure extended charge carrier lifetimes under oxidation: Time-domain ab initio analysis. *J. Phys. Chem. Lett.* **2019**, *10*(5), 1083–109.

(36) Komsa, H. P.; Kotakoski, J.; Kurasch, S.; Lehtinen, O.; Kaiser, U.; Krasheninnikov, A. V. Two-dimensional transition metal dichalcogenides under electron irradiation: defect production and doping. *Phys. Rev. Lett.*, **2012**, *109*, 035503.

(37) Kistanov, A. A.; Cai, Y.; Zhou, K.; Dmitriev, S. V.; Zhang, Y. W. The role of $H_2O$ and $O_2$ molecules and phosphorus vacancies in the structure instability of phosphorene. *2D Mater.* **2017**, *4*, 015010.

(38) Kistanov, A. A.; Nikitenko, V. R.; Prezhdo, O. V. Point defects in two-dimensional γ-phosphorus carbide. *J. Phys. Chem. Lett.* **2021**, *12*(1), 620–626.

(39) Hu, W.; Yang, J. Defects in phosphorene. *J. Phys. Chem. C* **2015**, *119*, 20474–20480.

(40) Wang, P.; Qiao, L.; Xu, J.; Li, W.; Liu, W. Erosion mechanism of $MoS_2$-based films exposed to atomic oxygen environments. *ACS Appl. Mater. Interfaces* **2015**, *7*(23), 12943–12950.

(41) Krasheninnikov, A. V.; Lehtinen, P. O.; Foster, A. S.; Nieminen, R. M. Bending the rules: contrasting vacancy energetics and migration in graphite and carbon nanotubes. *Chem. Phys. Lett.* **2006**, *418*, 132–136.

(42) Hong, J. Exploring atomic defects in molybdenum disulphide monolayers. *Nat. Commun.* **2015**, *6*, 6293.

(43) Chizallet, C.; Digne, M.; Arrouvel, C.; Raybaud, P.; Delbecq, F.; Costentin, G.; Che, M.; Sautet, P.; Toulhoat, H. Insights into the geometry, stability and vibrational properties of OH groups on γ-$Al_2O_3$, $TiO_2$-anatase and MgO from DFT calculations. *Top. Catal.* **2009**, *52*, 1005–1056.

(44) Sun, P.-P.; Kripalani, D. R.; Hao, M.; Chi, W.; Li, W.; Zhou K. Emissive nature and molecular behavior of zero-dimensional organic−inorganic metal halides $Bmpip_2MX_4$. *J. Phys. Chem. Lett.* **2020**, *11*(13), 5234−5240.

(45) Sun, P.-P.; Kripalani, D. R.; Chi, W.; Snyder, S. A.; Zhou, K. High carrier mobility and remarkable photovoltaic performance of two-dimensional Ruddlesden–Popper organic–inorganic metal halides $(PA)_2(MA)_2M_3I_{10}$ for perovskite solar cell applications. *Mater. Today* **2021**, *47*, 45−52.

(46) Shcherbinin, S. A.; Zhou, K.; Dmitriev, S. V.; Korznikova, E. A.; Davletshin, A. R.; Kistanov, A. A. Two-dimensional black phosphorus carbide: Rippling and formation of nanotubes. *J. Phys. Chem. C* **2020**, *124*(18), 10235–10243.

(47) Sun, P.-P.; Kripalani, D. R.; Bai, L.; Chi, W.; Zhou, K. Pentadiamond: A highly efficient electron transport layer for perovskite solar cells. *J. Phys. Chem. C* **2021**, *125*(9), 5372−5379.

(48) Frisenda, R.; Navarro-Moratalla, E.; Gant, P.; De Lara, D. P.; Jarillo-Herrero, P.; Gorbachev, R. V.; Castellanos-Gomez, A. Recent progress in the assembly of nanodevices and van der Waals heterostructures by deterministic placement of 2D materials. *Chem. Soc. Rev.* **2018**, *47*, 53–68.

(49) Kistanov, A. A.; Khadiullin, S. Kh.; Zhou, K.; Dmitriev, S. V.; Korznikova, E. A. Environmental stability of bismuthene: Oxidation mechanism and structural stability of 2D pnictogens. *J. Mater. Chem. C* **2019**, *7*, 9195–9202.

(50) Kistanov, A. A.; Cai, Y.; Zhou, K.; Dmitriev, S V.; Zhang, Y.-W. Atomic-scale mechanisms of defect- and light-induced oxidation and degradation of InSe. *J. Mater. Chem. C* **2018**, *6*, 518–525.

(51) Katin, K. P.; Prudkovskiy, V. S.; Maslov, M. M. Influence of methyl functional groups on the stability of cubane carbon cage. *Phys. E* **2016**, *81*, 1–6.

(52) Lloret, V.; Rivero-Crespo, M. A.; Vidal-Moya, J. A.; Wild, S.; Domenech-Carbo, A.; Heller, B. S. J.; Shin, S.; Steinruck, H. P.; Maier, F.; Hauke, F.; Varela, M.; Hirsch, A.; Leyva-Perez, A.; Abellan, G. Few layer 2D pnictogens catalyze the alkylation of soft nucleophiles with esters. *Nat. Commun.* **2019**, *10*, 509.

(53) Kistanov, A. A.; Korznikova, E. A.; Huttula, M.; Cao, W. The interaction of two-dimensional α- and β-phosphorus carbide with environmental molecules: A DFT study. *Phys. Chem. Chem. Phys.* **2020**, *22*, 11307–11313.

(54) Lin, S. H.; Kuo J. L. Towards the ionic limit of two-dimensional materials: Monolayer alkaline earth and transition metal halides. *Phys. Chem. Chem. Phys.* **2014**, *16*, 20763–20771.

(55) Botella, R.; Chiter, F.; Costa, D.; Nakashima, S.; Lefevre, G., Influence of hydration/dehydration on adsorbed molecules: Case of phthalate on goethite. *Colloids Surf. A Physicochem. Eng.* **2021**, *625*, 126872.

(56) Radisavljevic, B.; Radenovic, A.; Brivio, J.; Giacometti, V.; Kis A. Single-layer $MoS_2$ transistors. *Nat. Nanotech.* **2011**, *6*, 147–150.

(57) Kresse, G.; Furthmuller, J. Efficient iterative schemes for ab initio total-energy calculations using a plane-wave basis set. *Phys. Rev. B* **1996**, *54*, 11169.

(58) Perdew, J. P.; Burke, K.; Ernzerhof, M. Generalized gradient approximation made simple. *Phys. Rev. Lett.* **1996**, *77*, 3865−3868.

(59) Heyd, J.; Scuseria, G. E.; Ernzerhof, M. Hybrid functionals based on a screened coulomb potential. *J. Chem. Phys.* **2003**, *118*, 8207.

(60) Togo, A.; Tanaka, I. First-principles phonon calculations in materials science. *Scr. Mater.* **2015**, *108*, 1−5.

(61) Tersoff, J., Hamann, D. R. Theory of the scanning tunneling microscope. *Phys. Rev. B* **1985**, *31*, 805.

(62) Le Page, Y.; Saxe, P. Symmetry-general least-squares extraction of elastic data for strained materials from ab initio calculations of stress. *Phys. Rev. B: Condens. Matter Mater. Phys.* **2001**, *63*, 174103.

(63) Gaillac, R; Pullumbi, P and Coudert, F.-C. ELATE: An open-source online application for analysis and visualization of elastic tensors. *J. Phys.: Condens. Matter.* **2016**, *28*, 275201.